\documentclass{aa}

\begin{document}
\input{psfig}
\newbox\grsign \setbox\grsign=\hbox{$>$} \newdimen\grdimen \grdimen=\ht\grsign
\newbox\simlessbox \newbox\simgreatbox
\setbox\simgreatbox=\hbox{\raise.5ex\hbox{$>$}\llap
     {\lower.5ex\hbox{$\sim$}}}\ht1=\grdimen\dp1=0pt
\setbox\simlessbox=\hbox{\raise.5ex\hbox{$<$}\llap
     {\lower.5ex\hbox{$\sim$}}}\ht2=\grdimen\dp2=0pt
\def\simgreat{\mathrel{\copy\simgreatbox}}
\def\simless{\mathrel{\copy\simlessbox}}
\newbox\simppropto
\setbox\simppropto=\hbox{\raise.5ex\hbox{$\sim$}\llap
     {\lower.5ex\hbox{$\propto$}}}\ht2=\grdimen\dp2=0pt
\def\simpropto{\mathrel{\copy\simppropto}}

\title{The Metal Content of the Bulge Globular Cluster NGC 6528
\thanks{Observations collected both at the European Southern Observatory, 
Paranal and La Silla, Chile (ESO programme 65.L-0340) and with the 
NASA/ESA Hubble Space Telescope, obtained at the Space Telescope Science 
Institute, operated by AURA Inc. under contract to NASA.
} }

\author{
M. Zoccali\inst{1,2}
\and
B. Barbuy\inst{3}
\and 
V. Hill\inst{4}
\and
S. Ortolani\inst{5}
\and
A. Renzini\inst{6}
\and
E. Bica\inst{7}
\and
Y. Momany\inst{5}
\and
L. Pasquini\inst{6}
\and
D. Minniti\inst{1}
\and
R.M. Rich\inst{8}
}
\offprints{M. Zoccali}
\institute{
Universidad Catolica de Chile, Department of Astronomy \& Astrophysics,
Casilla 306, Santiago 22, Chile;\\
e-mail: mzoccali@astro.puc.cl, dante@astro.puc.cl
\and
Princeton University Observatory, Peyton Hall, Princeton, NJ 08544, USA;
\and
Universidade de S\~ao Paulo, IAG, Rua do Mat\~ao 1226,
Cidade Universit\'aria, S\~ao Paulo 05508-900, Brazil;\\
e-mail: barbuy@astro.iag.usp.br
\and
Observatoire de Paris-Meudon,  92195 Meudon Cedex, France;\\
e-mail: Vanessa.Hill@obspm.fr
\and
Universit\`a di Padova, Dipartimento di Astronomia, Vicolo
 dell'Osservatorio 2, I-35122 Padova, Italy;\\
 e-mail: ortolani@pd.astro.it, momany@pd.astro.it
\and
European Southern Observatory, Karl Schwarzschild Strasse 2, 85748
Garching bei M\"unchen, Germany;\\
e-mail:  arenzini@eso.org, lpasquin@eso.org
\and
Universidade Federal do Rio Grande do Sul, Departamento de Astronomia, 
CP 15051, Porto Alegre 91501-970, Brazil;\\
e-mail: bica@if.ufrgs.br
\and 
UCLA, Department of Physics \& Astronomy, 8979 Math-Sciences Building,
Los Angeles, CA 90095-1562, USA;\\
e-mail: rmr@astro.ucla.edu
}

\date{Received ; accepted }

\abstract{
High resolution spectra of five  stars  in the bulge globular  cluster
NGC 6528  were obtained  at the 8m  VLT UT2-Kueyen  telescope with the
UVES spectrograph. Out of the five stars,  two of them showed evidence
of binarity.  The target stars belong to  the horizontal and red giant
branch stages,   at  4000 $<$  T$_{\rm eff}$  $<$  4800 K.   Multiband
$V,I,J,H,K_{\rm  s}$ photometry was used  to  derive initial effective
temperatures and gravities.  The  main purpose of   this study is  the
determination  of metallicity and   elemental ratios for this template
bulge  cluster, as    a  basis for   the  fundamental calibration   of
metal-rich populations.   The present  analysis provides a metallicity
[Fe/H] =  $-0.1\pm0.2$ and the  $\alpha$-elements  O, Mg and  Si, show
[$\alpha$/Fe] $\approx$ +0.1, whereas Ca  and Ti are around the  solar
value  or below, resulting in   an  overall  metallicity Z   $\approx$
Z$_{\odot}$.
}
\titlerunning{Metal content of NGC 6528}
\authorrunning{M. Zoccali et al.}
\maketitle

\section{Introduction} 

The globular cluster NGC~6528 is located in Baade's Window, at (J2000)
$\alpha=18^{\rm h}04^{\rm m}51.5^{\rm s}$, $\delta=-30^{\rm o}03'04''$
(l=1.1459$^{\circ}$,   b=$-4.1717^{\circ}$),  at       a      distance
d$_{\odot}=7.8$ kpc  from the Sun and  ${\rm R_{GC}}=0.6$ kpc from the
Galactic center (Barbuy et al. 1998).

Ortolani et al.   (1992)  first presented BVRI    CCD Colour-Magnitude
Diagrams (CMD)  of NGC 6528.  Ortolani et  al.  (1995) showed that NGC
6528, together with its  ``twin''  NGC 6553,  has a  Colour  Magnitude
Diagram (CMD)  and a luminosity function  very similar  to that of the
Galactic bulge, as  seen in the Baade's Window  field. This is a clear
indication   that  the  two   populations  have  comparable  age   and
metallicity.

For this reason,  the two bulge  globular clusters (GCs) NGC 6528  and
NGC 6553 have  often been used  as  templates for the  old, metal-rich
population of  the Milky Way  central spheroid  (e.g.  Zoccali  et al.
2003).

Despite its key importance for the interpretation  of the formation of
our own Galaxy and, by extension, of  external spheroids, the absolute
value of the Fe and $\alpha$-element abundance, for  both NGC 6528 and
NGC 6553 still are a matter of debate.  The main reasons for this are:
{\it i)} the intrinsic  difficulty to observe  these faint stars; {\it
ii)} the  ambiguous location of the continuum,  due to the severe line
crowding  in  these metal-rich  objects,   {\it iii)}  the presence of
strong molecular (mainly TiO) bands  in the brightest, coolest  stars,
together   with  effects  of  $\alpha$-element   enhancements   on the
continuum absorption;  {\it  iv)} the  uncertainty of   the  effective
temperature scale   for stars as  hot as  the  Horizontal Branch (HB):
temperatures derived by imposing excitation equilibrium for the
\ion{Fe}{I} lines tend to be overestimated probably due to blends, and
possibly to NLTE effects on \ion{Fe}{I} in giants, whereas
temperatures derived from colours show uncertainties due to reddening
variations.

A few stars in each of the two GCs have been  observed in the past few
years at high  spectral   resolution, but derived   metallicities show
discrepancies,  higher than the  estimated uncertainty, thus revealing
that there are systematic errors not properly taken into account.

In NGC~6553, two were observed with CASPEC, at the ESO 3.6m Telescope,
by    Barbuy  et   al.      (1999).   They    derived  [Fe/H]=$-0.55$,
[Na/Fe]$\approx$[Al/Fe]$\approx$[Ti/Fe]$\approx$+0.5,
[Mg/Fe]$\approx$[Si/Fe]$\approx$[Ca/Fe]$\approx$+0.3               and
[s-elements/Fe]$\simless$0.0.   A significantly higher metallicity was
obtained   instead  by   Cohen  et   al.  (1999),   from  HIRES@Keck~I
observations   of          5   HB          stars:      [Fe/H]=$-0.16$,
[$\alpha$/Fe]$\approx$+0.2.  More   recently,  Origlia et  al.  (2002)
measured the  metallicity of  NGC~6553 from  near-IR  spectra obtained
with   NIRSPEC@Keck~II,       obtaining       [Fe/H]=$-0.3$,      with
[$\alpha$/Fe]=+0.3. Mel\'endez et al. (2003) used PHOENIX@Gemini-South
to obtain spectra of R = 50 000 in the H band, and they derived [Fe/H]
= -0.20 and  [O/Fe] = +0.20.   A  summary of the previous  metallicity
determinations   of   NGC~6553 is given    in  Table~5  of   Barbuy et
al. (1999).

For  NGC~6528 the only high-resolution  analysis  published so far has
been that of Carretta et al.  (2001; hereafter C01), who observed four
HB   stars,  with  HIRES   at Keck~I.    They  obtained
[Fe/H]=+0.07 and  [M/H] = +0.17.
Recently, Momany et al.  (2003) determined the metallicity of NGC~6528
from RGB morphology indicators, in a ($K$,$V-K$) CMD, obtaining values
in    the range -0.43 $<$ [Fe/H] $<$ 0.13    depending on  the  adopted
calibration.

In this work we present detailed abundance  analysis of three stars in
NGC 6528 using high resolution \'echelle spectra obtained with UVES at
the  ESO VLT-UT2  Kueyen  8m telescope, in  an  attempt  to reduce the
discrepancy   between previous    measurements using  higher  spectral
resolution   (R$\sim$50  000) relative    to  previous  studies  (e.g.
R$\sim$37  000 in C01  and R$\sim$20 000 in  Barbuy et al. 1999).  The
analysis is restricted  to relatively hot  (T$_{\rm eff}$ $>$ 4000  K)
stars, in  order to avoid the above  mentioned problems with molecular
bands.

The  paper  is  organized as follows.   Spectroscopic observations are
described in  Sect. 2. Sect.   3 focuses  on the  determination of the
stellar parameters: effective temperature  and gravity from photometry
and  spectroscopy.   Equivalent width and  oscillator   strenghts  are
discussed  in  Sect. 3.  Iron  abundance is  derived in   Sect.  4 and
abundance ratios in Sect. 5. The results are then discussed in Sect. 6.

\section {The Data} 

\subsection{Imaging}

$V$ and $I$ photometry of the central region  of NGC~6528 was obtained
using two sets  of WFPC2 observations with  the Hubble Space Telescope
(HST), (GO5436:  Ortolani   et al.   1995, and GO8696:    Feltzing and
Johnson  2002).  Infrared $J, H,  K_s$ observations were obtained with
the SOFI  infrared   camera   of  the ESO  New   Technology  Telescope
(NTT). The details  about both optical and  near-IR data are  given in
Momany et al. (2003).

\begin{figure}[ht]
\psfig{file=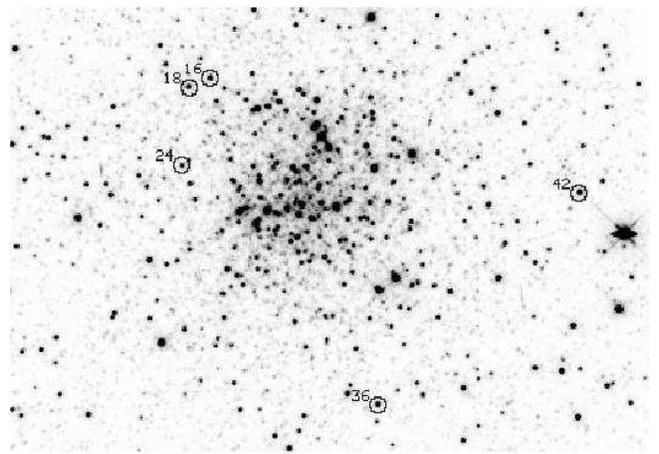,angle=0,width=8.5 cm}
\caption {Finding chart for the 5 stars observed in NGC 6528
(HST WFPC2 F814W image). Resolution is 0.1''/px.}
\label{chart}
\end{figure}

\begin{figure}[ht]
\psfig{file=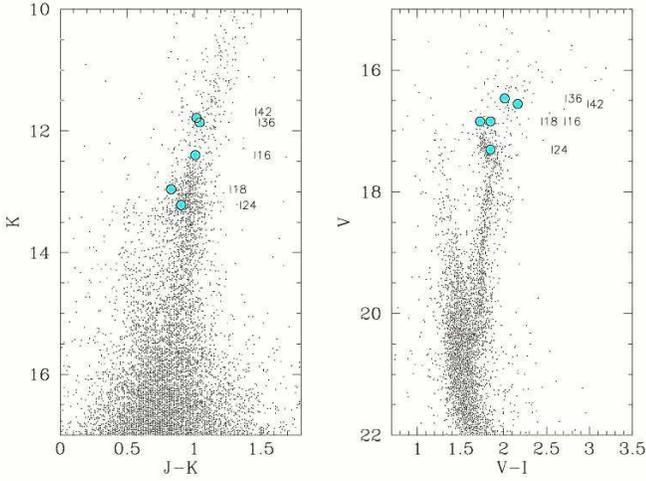,angle=-90,width=9.0 cm}
\caption {The SOFI near-IR (left) and HST optical (right) CMD of NGC~6528
with the target stars shown as big filled circles.}
\label{cmd}
\end{figure}

Five isolated  stars were selected  from our  sample for spectroscopic
observations (Fig.~\ref{chart}).  The target location on both the SOFI
near-IR and the HST optical CMD is shown in Fig.~\ref{cmd}.

\begin{figure}[ht]
\psfig{file=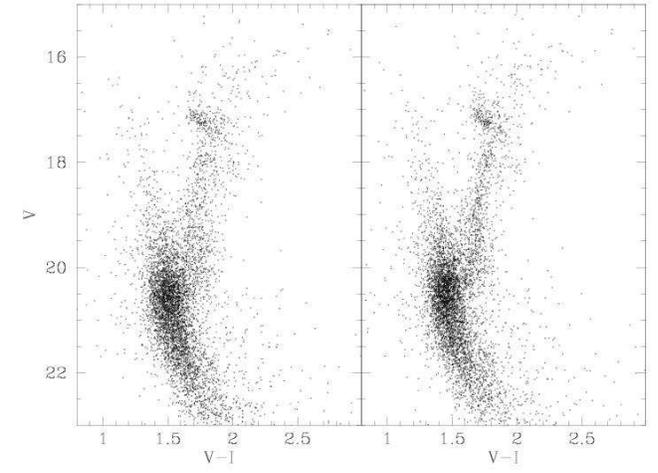,angle=-90,width=9.0 cm}
\caption {NGC 6528: Original and reddening corrected V vs. V-I
Colour Magnitude Diagrams}
\label{redd}
\end{figure}

Table~\ref{targets} lists the  coordinates, magnitudes and colours  of
the  sample stars.  The star's identifications  follow the notation by
van den Bergh \& Younger (1979), where the prefix I corresponds to the
internal  ring.   For  each star  the  first line  corresponds to  the
original   magnitudes, while  the  second   line  corresponds to   the
magnitudes  corrected  for  total (see   Sect.   3.1) and differential
extinction.  The latter correction has  been carried out following the
method described in Piotto et al.  (1999) and  Zoccali et al.  (2001).
This method assumes  that the shift  relative to the RGB fiducial mean
locus, between    different subregions, is  dominated  by differential
reddening.  The observed field is  divided into small subregions,  and
the reddening  of each  of them  is estimated  by comparison  with the
fiducial mean  locus  of NGC 6528  from  Momany  et  al.  (2003).  The
result  of this  procedure   for  the ($V$,$V-I$)    CMD is  shown  in
Fig.~\ref{redd}.

\begin{table*}
\caption[1]{Positions and magnitudes from Momany et al. (2003). For
each  star,  the first line lists  the  original  magnitudes while the
second gives dereddened  magnitudes and colours adopting the reddening
law by Dean  et al. (1978)  and  Rieke \&  Lebofsky (1985).  Star  I36
could not be measured in the  H frame. $(V-I)_J$ is Johnson, $(V-I)_C$
is Cousins.}
\begin{flushleft}
\tabcolsep 0.15cm
\begin{tabular}{cccccccccccccc}                                               
\noalign{\smallskip}
\hline
\noalign{\smallskip}
{\rm star} & $\alpha_{2000}$ & $\delta_{2000}$ & $\Delta$E($B-V$) & $V$ & $I$ & $J$ & $H$ & $K_{\rm s}$ &   &   &   &  \cr
&& &E(B-V) & $V_0$ & $I_0$ & $J_0$ & $H_0$ & $K_{\rm s0}$ & $(V-I)_{C0}$ & $(V-I)_{J0}$ & $(V-K)_0$  & $(J-K)_0$  \cr
\noalign{\vskip 0.2cm}
\noalign{\hrule\vskip 0.2cm}
\noalign{\vskip 0.2cm}
I-16 & 18:04:49.5 & -30:03:03.5 & -0.039      & 16.73 & 14.94 & 13.41 & 12.57 & 12.40 &      &      &      &      \cr 
     &            &             & 0.421 & 15.42  & 14.19 & 13.12 & 12.34 & 12.25 & 1.23 & 1.58 & 3.17 & 0.87 & \cr
I-18 & 18:04:49.7 & -30:03:02.5 & -0.039  & 16.73 & 15.06 & 13.79 & 13.03 & 12.96 &      &      &      &      \cr
     &            &             & 0.421 & 15.42 & 14.31 & 13.50 & 12.80 & 12.81 &  1.11 &  1.43 &  2.61 &  0.69 & \cr
I-24 & 18:04:50.3 & -30:03:08.9 & -0.039 & 17.19 & 15.40 & 14.12 & 13.39 & 13.22 &      &      &      &      \cr
     &            &             & 0.421 & 15.88 & 14.65 & 13.83 & 13.16 & 13.07 &  1.23  & 1.58 &  2.81  & 0.76 & \cr
I-36 & 18:04:50.9 & -30:03:48.0 & -0.026  & 16.39 & 14.42 & 12.90 & --    & 11.86 &      &      &      &      \cr
     &            &             & 0.434 & 15.04 & 13.65 & 12.59 & 99.75 & 11.71  & 1.39 &  1.79 &  3.34 &  0.89 & \cr

I-42 & 18:04:47.7 & -30:03:46.5 & -0.046  & 16.43 & 14.33 & 12.81 & 11.90 & 11.79 &      &      &      &      \cr 
     &            &             & 0.414 & 15.15 & 13.60 & 12.53 & 11.68 & 11.64  & 1.55 &  1.99 &  3.50 &  0.89 & \cr
\noalign{\smallskip} \hline \end{tabular}
\label{targets}
\end{flushleft} 
\label{starmag}
\end{table*}

\begin{figure}[ht]
\psfig{file=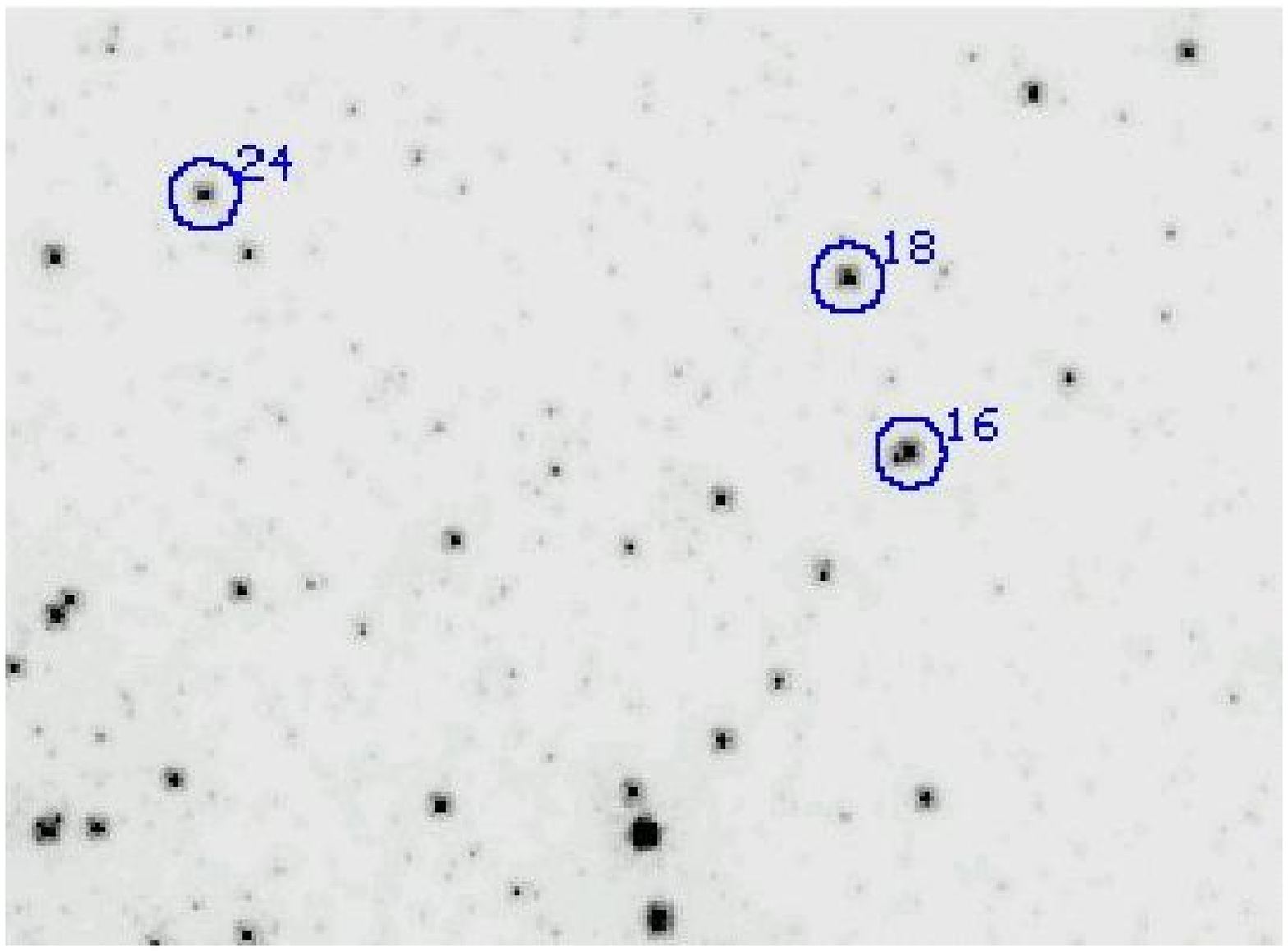,angle=0,width=8.5 cm}
\caption {Three program stars I16, I18 and I24 identified in a
NIC2 $J$ frame. Resolution is 0.075''/px.   The diameter of the circle
is 1 arcsec, corresponding to the size of the  UVES slit. Star I16 clearly
has a nearby  companion, not resolved  in the ground-based (nor in
the WFPC2) images.}
\label{nicmos}
\end{figure}

A posteriori, we  identified some of our  targets also in the field of
view of a  $J,H$ set of NICMOS@HST  data (Ortolani et  al. 2001).  The
spatial resolution of the NICMOS  observations allowed us to  discover
that  I-16  has in fact  a  nearby  companion which  contaminates  the
spectrum by about 23$\%$ in the J and H bands (see  Fig.~\ref{nicmos}). 
For this reason,
I-16 was discarded from the spectroscopic analysis.

\subsection{Spectra}

High  resolution   spectra for  five  members  of   NGC~6528,  in  the
wavelength range  $\lambda\lambda$ 4800-6800  {\rm \AA}, were obtained
with the UVES spectrograph at the  ESO VLT.  The target membership had
been  previously   verified  from   radial velocities  by    Coelho et
al.  (2001).  In what  follows  we will  consider  mainly  the reddest
portion of  the  spectrum (5800-6800  {\rm  \AA}) covered   by the MIT
backside illuminated and AR coated CCD ESO  \# 20 of 4096x2048 pixels,
of pixel  size 15x15$\mu$m.  The bluer  region was used only to derive
the carbon  abundance from the C$_2$ (0,1)  bandhead at $\lambda$ 5635
{\rm \AA}.  With   the UVES   standard   setup   580, the   resolution   
is R$\sim45 000$ for a  1 arcsec slit  width, and  R$\sim55 000$ for a
slit of  0.8    arcsec.  The pixel   scale is    0.0174  {\rm \AA}/px.
The log of the observations is shown in Table~\ref{logobs}.

\begin{table*}
\caption[2]{Log of the spectroscopic observations carried out on 2000 
June 24-25-26 and 2001 July 6. The quoted seeing is the mean value
along the exposures.}
\begin{flushleft}
\begin{tabular}{llllllllccc}
\noalign{\smallskip}
\hline
\noalign{\smallskip}
Target & V & date/ & UT & exp & Seeing  & Airmass & (S/N)/px  & Slit & ${\rm v_r^{obs}}$ & 
${\rm v_r^{hel.}}$ \\
\noalign{\smallskip}
 & & Julian day &  & (s) & ($''$) & & & width &${\rm km s^{-1}}$ &${\rm km s^{-1}}$  
 \\
\noalign{\smallskip}
\noalign{\smallskip}
\hline  
\noalign{\vskip 0.2cm}
%
I-16 & 17.496 & 26.06.00 & 03:46:28 & 5400 & 1.0 & 1.0     &    & 1.0$''$ & 208 & 206 \\
     &        & 2451721  & 05:18:53 & 5400 & 0.8 & 1.0-1.2 &    &  ''     & ''  & ''  \\
I-18 & 16.732 & 26.06.00 & 03:46:28 & 5400 & 1.0 & 1.0     & 40   & 1.0$''$ & 211 & 209 \\
     &        & 2451721  & 05:18:53 & 5400 & 0.8 & 1.0-1.2 &    &   ''    & ''  & ''  \\
I-24 & 17.194 & 26.06.00 & 06:59:18 & 4584 & 1.1 & 1.2-1.6 & 30   & 1.0$''$ & 215 & 213 \\
     &        & 2451721  & 08:37:43 & 3600 & 1.3 & 1.7-2.6 &    &   ''    & ''  & ''  \\
I-36 & 16.392 & 26.06.00 & 00:13:03 & 5400 & 1.1 & 1.2-1.8 & 40   & 1.0$''$ & 218 & 217 \\
     &        & 2451721  & 01:48:35 & 5400 & 1.1 & 1.2-1.0 &    &    ''   & ''  & ''  \\
I-42 & 16.426 & 06.07.01 & 05:29:42 & 3600 & 0.7 & 1.0-1.1 & 30   & 0.8$''$ & 222 & 215 \\
      &       & 2452096  & 06:30:35 & 3600 & 0.8 & 1.2-1.5 &    &    ''   & ''  & ''  \\
\noalign{\smallskip}
\hline
\end{tabular}
\end{flushleft} 
\label{logobs}
\end{table*}

The spectra were reduced using the UVES context of the MIDAS reduction
package, including  bias   and   inter-order background   subtraction,
flatfield correction, extraction and wavelength calibration (Ballester
et al.  2000).  When possible we used  the optimal extraction routine,
rejecting the cosmic rays. However, in a few cases (e.g.  star I16 and
I18,    which were observed  through   the   same  slit), the  optimal
extraction could not be used, and therefore we  turned to the standard
extraction routine. The two spectra obtained for each target star were
summed before the analysis.

A mean v$_{\rm  r}$  = 215$\pm$5  km  s$^{\rm {-1}}$  or  heliocentric
v$^{\rm hel}_{\rm r}$ = 212 km s$^{\rm {-1}}$  was found for NGC~6528,
higher than the values of 160 kms $^{\rm {-1}}$ given  in Zinn \& West
(1984) or 143 km s$^{\rm {-1}}$ given in Zinn (1985), but in very good
agreement with more recent values of $203\pm20$  km s$^{\rm {-1}}$ and
212$\pm$16 km s$^{\rm {-1}}$  measured respectively by Minniti  (1995)
and Rutledge et al. (1997).

\section{Stellar Parameters}

\subsection{Reddening}

A reddening of  E($B-V$)  = 0.55 was  given in  Momany et al.  (2003),
however this was a mean value from determinations relative to NGC 6553
and 47 Tucanae,  in different colours  (see their Table~3),  and those
values are somewhat  dependent on the metallicity and $\alpha$-element
assumptions in isochrone models. They can  be used as guidelines, but
for temperature  determinations we need  to  investigate further about
the value  to  adopt,  since a  $\Delta$E(B-V)  = 0.07  implies  in  a
$\Delta$T$_{\rm eff}$ $\approx$ 400 K.

{\it Reddening vs. spectral type:}

A very well established   result, after the analysis of  Schmidt-Kaler
(1961), is that the absorption (in the $UBV$ bands) is a function of the
the energy distribution of the star within  the broadband filters (see
also more recent discussions  and  calculations in Olson  1975, Grebel
and Roberts  1995, and Fitzpatrick 1999).   Such effect is  due to the
decrease of the extinction, toward longer wavelengths, across the width
of  the filter.  The net effect  results in  a  shift of the effective
wavelength, for  a reddened star, to a  longer wavelength than  for an
unreddened  star. The  magnitude of  the shift  depends on  the energy
distribution (therefore on the spectral type) of  the star. This makes
redder (cooler) stars   systematically less absorbed.   The  effect is
more  pronounced in the  B  band than  in  V, for  the same amount  of
interstellar dust, and thus the colour excess  becomes sensitive to the
stellar   energy  distribution.   Schmidt-Kaler  (1961)  introduced  a
parameter $\eta$ to quantify this effect:

$\eta$ = E(B-V)$_{\rm (Sp T)}$/E(B-V)$_{\rm B0}$ 

\noindent      
and showed that its variation is of the order of 10\% from early type
stars to M stars. $\eta$ is also slightly dependent on the reddening
itself.

NGC 6522 is  another globular cluster located   in the Baade's  Window
that can be used  to derive colour excess  in the region.  The work by
Terndrup \&    Walker  (1994) on  several fields    in NGC 6522  gives
E(B-V)=0.52 for B-V=0 stars,  corresponding to about 0.44-0.45 for our
K-M stars.  This is done  via a simple fit to  CMDs  relative to other
globular  clusters having the same    RGB morphology. Terndrup et  al.
(1998) derived Av=1.4 on a proper motion cleaned CMD of NGC 6522. This
gives  again  about E(B-V)=0.45 for K-M   stars. We can  conclude that
there is quite strong evidence that NGC 6522  has a reddening close to
0.44-0.45.  Now we should estimate the difference between NGC 6528 and
NGC 6522. A  check  on Stanek's  (1996)  maps indicates that  the  two
regions  have   very similar extinctions.   The   zero point  has been
revised by Alcock et al.  (1998) and Gould  et al.  (1998).  Following
Alcock et al., from  their Table~1, the extinction  near NGC  6528 (at
$\alpha$=18$^{\rm      h}$04$^{\rm          m}$29.46$^{\rm        s}$,
$\delta$=-30$^{\circ}$01'  07'', J2000.0)  is $\Delta$A$_{\rm V}$=0.05
mag higher than near NGC 6522. Using a total-to-selective absorption R
= A$_{\rm  V}$/E(B-V) = 3.1, this transforms  into E(B-V)= 0.01 - 0.02
mag,  which makes a very small   difference.  The same authors derived
E(B-V)=0.48  for (B-V)=0.0 stars  for  the less  redenned side of  the
Baade Window, corresponding to one   of the lowest values (about  0.42
for K-M stars).  They strongly  argue against high values.  Since they
start from luminosity measurements of the RR Lyrae stars, they have to
make assumptions about the total-to-selective absorption R$_{\rm V}$ =
A$_{\rm V}$/E(B-V) value to get the E(B-V), but  still this appears to
be a quite firm result.
  
We carried out an independent reddening measurement for NGC 6522 using
HST photometry from  Piotto et  al. (2002).   The CMD of  NGC 6522 was
compared  to   those    of  reference clusters,     having  comparable
metallicities, and which  have been  observed  in  with the same   HST
instrumentation. The reddening   difference between NGC 6522 and  each
reference cluster is measured  as the colour  difference of their  red
giant branches, $\Delta$(B-V)$_{\rm RGB}$=$\Delta$ E(B-V), as  shown in
Table~\ref{tabredd}.

A cluster metallicity can be estimated  from the curvature of the RGB,
since more metal-rich giants show progressively fainter RGB magnitudes
(as well as redder RGB colours), due to higher blanketing (Ortolani et
al. 1990, 1991).  The metallicities reported are from Harris (1996, as
updated                                                             in
http://phy\-sun.\-phy\-sics.mc\-mas\-ter.\-ca/Glo\-bu\-lar.\-html).
Note that the more metal rich clusters will give obviously a bit lower
reddening and viceversa: M 13, NGC 6681 and NGC 1904  appear to have a
more metal  poor morphology  as  concerning the RGB,  and give  higher
$\Delta$(B-V). The E(B-V) = $\Delta$(B-V)  would be given by comparing
two  clusters  of  same metallicity,   given that  there should  be no
blanketing differences between them.  The  best morphological fits are
obtained  for NGC 5904 (M5), NGC  1851 and NGC  6266.   Harris (1996) gives
[Fe/H]=-1.44 for    NGC    6522, and  in   these   cases  the  derived
$\Delta$(B-V) is  close to the  E(B-V) value.  In conclusion, it seems
that E(B-V)=0.42-0.50 represents a fiducial  reddening value for NGC
6522 for the colour of horizontal branch to G-K stars.

In order to obtain the reddening for NGC 6528 we would have to add the
reddening  difference  between the  former and  NGC 6522  (+0.02), but
another $\Delta$E(B-V)$=-0.02$ would  be needed to translate this  for
K-M  stars (c.f. parameter $\eta$ defined   above). Therefore, for our
stars, we adopt  the  same reddening   E(B-V)=$0.46\pm0.04$ determined
above for NGC 6522. Note  that this result  is independent of possible
zero point errors in the Piotto et  al. photometry because it is based
on differential photometry.

Yet another free parameter comes into play, because the extinction law
towards the  bulge is somewhat  controversial. Extremely low values of
the  total-to-selective  absorption R$_{\rm V}$   = A$_{\rm V}$/E(B-V)
were  recently proposed by e.g.   Udalski (2003) in the galactic bulge
OGLE II  field, as  opposed to  the traditional  R$_{\rm V}$=3.1  from
Cardelli et al. (1989) or Rieke \& Lebofsky (1985).  For completeness,
let  us recall that    the total-to-selective absorption  R$_{\rm  V}$
depends on the  amount  of reddening  (Olson 1975)  and on metallicity
(Grebel  \& Roberts 1995),  as discussed in  Barbuy et al. (1998), and
the  value  should be somewhat   higher  for high reddening  and  high
metallicity  values.  If lower   R$_{\rm V}$ values  are  assumed, our
resulting dereddened colours will be redder, and therefore the deduced
effective temperatures will be cooler.

For the present analysis we  adopt E(B-V) = 0.46  for NGC 6528 and the
extinction law given  by  Dean et al.   (1978) and  Rieke  \& Lebofsky
(1985),     namely:  R$_{\rm  V}$   =    3.1,  E($V-I$)/E($B-V$)=1.33,
E($V-K$)/E($B-V$)=2.744 and E($J-K$)/E($B-V$)=0.527 (Table~\ref{targets}).

\begin{table}
\caption[1]{Reference clusters, respective metallicities
(according to Harris (1996)), and $\Delta$(B-V) corresponding
to the reddening of NGC 6522 derived differentially relative to the
comparison clusters.
}
\begin{flushleft}
\begin{tabular}{lllrrr}
\noalign{\smallskip}
\hline
\noalign{\smallskip}
{\rm comparison cluster} & [Fe/H] & {\rm $\Delta$(B-V)} \\
\noalign{\smallskip}
\hline
\noalign{\smallskip}
        M 13   &                -1.54  &           0.60 \\
        NGC 6229   &                -1.43   &           0.50 \\
        NGC 6723   &                -1.12  &           0.50 \\
        NGC 6681   &                -1.51  &           0.60 \\
        NGC 6266   &                -1.29  &           0.42 \\
        NGC 5904   &                -1.27  &           0.50 \\
        NGC 1851   &                -1.22  &           0.47 \\
        NGC 1904   &                -1.57  &           0.55 \\
\noalign{\smallskip} \hline \end{tabular}
\end{flushleft} 
\label{tabredd}
\end{table}

\begin{table*}
\caption[1]{
Stellar parameters derived using the calibrations by Alonso et al. (1999; AAM99) 
for $V-I$, $V-K$, $J-K$. The dereddened colours were obtained using the
reddening law by Dean et al. (1978) and Rieke \& Lebofsky (1985).}
\begin{flushleft}
\begin{tabular}{lllllllrrr}
\noalign{\smallskip}
\hline
\noalign{\smallskip}
& \multispan5 AAM99     &&&\\
\cline{2-6}     \\
{\rm star} & 
T$_{\rm eff}$(V-I) & 
T$_{\rm eff}$(V-K) & 
T$_{\rm eff}$(J-K) & 
T$_{\rm eff}$ (mean) & 
${\rm BC_{V}}$ & 
${\rm M_{bol}}$ & 
log g \\
\noalign{\vskip 0.2cm}
\noalign{\hrule\vskip 0.2cm}
\noalign{\vskip 0.2cm}
I-18 & 4504 & 4520 & 4409 & 4511 & --0.47 & --0.16 & 2.0 \\
I-36 & 4071 & 4051 & 3937 & 4019 & --0.87 & --0.93 & 1.5 \\
I-42 & 3897 & 3975 & 3937 & 3936 & --0.97 & --0.93 & 1.5 \\
\noalign{\smallskip} \hline \end{tabular}
\end{flushleft} 
\label{tabteff}
\end{table*}

\subsection{Temperatures}

\begin{figure}[ht]
\psfig{file=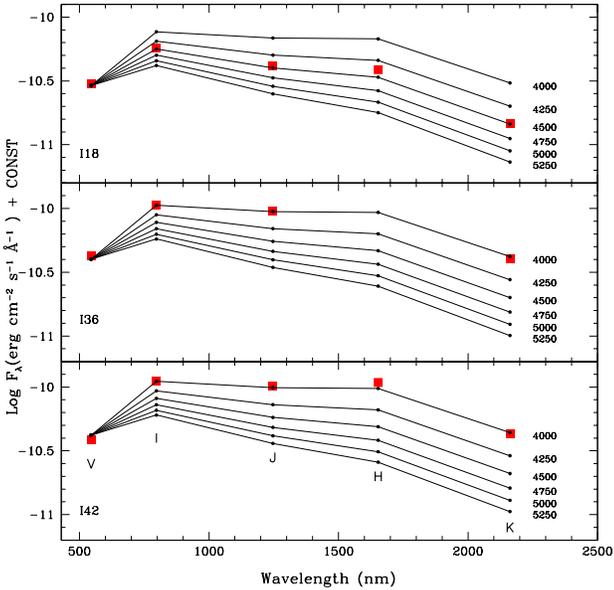,angle=0,width=8.5 cm}
\caption {Calibrations of Alonso et al. (1999), of $V$, $I$, $J$, $H$
 and $K$
as a function of effective temperature (solid lines), where the
magnitudes for  the sample stars are overplotted (filled squares).}
\label{SED}
\end{figure}

Effective temperatures were derived from  $V-K$, $V-I$ and $J-K$ using
the colour-temperature calibrations  by Alonso et al.  (1999) together
with the reddening  laws by Dean  et al. (1978)  and Rieke \& Lebofsky
(1985). The T$_{\rm eff}$  derived   from each  colour are  listed  in
Table~\ref{tabteff}.  We have excluded from the spectroscopic analysis
the star I-16 as explained in  Sect. 2.1, and  also star I-24, being a
suspect unresolved binary, for the reasons  explained below (Sect. 4).
In  order  to verify that   all the colours    are compatible with the
adopted  temperature, we  compared  the  spectral energy  distribution
(SED) of  our targets  with the prediction  by  Alonso et al.  (1999).
Fig.~\ref{SED}  shows the  flux  obtained from  the $V,I,J,H$  and $K$
magnitudes, together with Alonso  et  al.'s calibrations for stars  of
different temperatures  (4000-5250 K).   Since  Alonso et al.   give a
relation between colours and temperatures,  the predicted flux in each
band was  obtained   by  assuming an   arbitrary  value  for  the  $V$
magnitude.  A vertical shift was then allowed to match the SEDs of the
sample stars.

\subsection {Gravities}

The classical  relation  $\log g_*=  4.44   + 4\log (T_*/T_{\odot})  +
0.4(M_{\rm bol}-4.75) +  \log   (M_*/M_{\odot}$) was  used, adopting
T$_{\odot}$ = 5770 K, M$_*$ = 0.85 M$_{\odot}$ and M$_{\rm bol \odot}$
= 4.75 (Cram 1999).

A distance modulus of (m-M)$_0$ = 15.10 was adopted from Harris (1996)
together  with a reddening   of E($B-V$) = 0.46   and A$_V$ = 1.43, as
discussed  in the  previous sections. The  bolometric corrections from
Alonso et al.   (1999)   and  corresponding gravities are   given   in
Table~\ref{tabteff}.

Spectroscopic  gravities  were  also  derived by  imposing  ionization
equilibrium between FeI and FeII lines (see Sect. 4).


\section{Equivalent widths and oscillator strengths}

\begin{figure}[ht]
\psfig{file=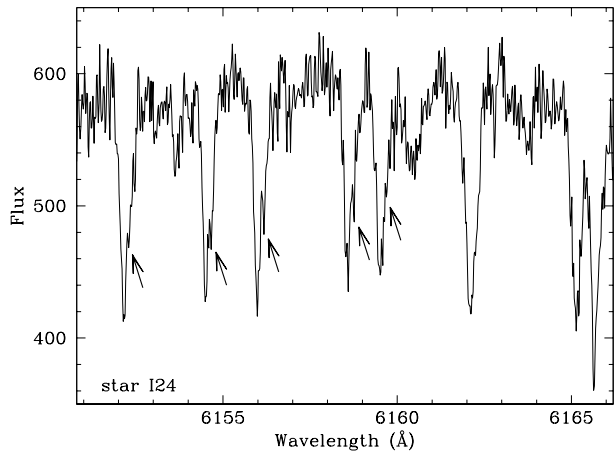,angle=-90,width=8.5 cm}
\caption{Spectrum of I24 showing its probable binary nature.}
\label{i24}
\end{figure}

The  equivalent widths have been measured  using a  new automatic code
developed   by P.   Stetson    (DAOSPEC,  Stetson et  al.  2004,    in
preparation).  This routine fits a Gaussian profile  to all lines,
imposing  a constant FWHM. In order  to  take into account the broader
profile of strong lines, we manually measured  lines stronger than 140
{\rm m\AA} allowing a best  fit for the  FWHM of each line.  Star I-24
shows lines  with FWHM significantly  larger  than that  of the  other
stars (a factor  of 1.5),  and most  of  the lines have  a  systematic
asymmetry (Fig.~\ref{i24}). This  was interpreted as  evidence of the
presence of an unresolved companion, and for this  reason the star was
excluded  from the  following analysis.  Note   that at lower spectral
resolution such unresolved binaries would  not have been detected, and
therefore the results  would  not  have  reflected the true    stellar
abundances.

The \ion{Fe}{I} line list and respective oscillator strengths given in
NIST (Martin et al. 2002) and some lines with gf-values from McWilliam
\&  Rich  (1994,  hereafter MR94)  were  used to  derive spectroscopic
parameters. Eight measurable \ion{Fe}{II}  lines, and their respective
oscillator strengths from Bi\'emont et al. (1991), and renormalized by
Mel\'endez \& Barbuy    (2004, in prepararion),    
were used to     obtain  ionization
equilibrium.

The damping constants were  computed where possible, and in particular
for most of the  \ion{Fe}{I}  lines, using the collisional  broadening
theory of  Barklem  et al. (1998,  2000  and references therein).  For
lines of Si, Ca and Ti the derived  values of the interaction constant
C$_6$, where C$_6$  =  6.46E-34 ($\sigma$/63.65)$^{5/2}$,  $\sigma$ being
the cross section of  collision computed by  Barklem et al., which is
related    to     the collisional    damping     constant $\gamma_6$ =
17v$^{3/5}$C$_6^{2/5}$N,  are reported   in Table \ref{tablines}.   
Otherwise a fit to the solar spectrum or standard van der Waals values
(Uns\"old 1955; Gray 1976) are employed.

The oscillator strengths of Na, Mg, Si, Ca and Ti lines given by Brown
\& Wallerstein  (1992), MR94, the NIST  database (Martin et al. 2002),
Barbuy  et  al.  (1999)  from  fits to   the solar  spectrum, the VALD
database (Kupka et al. 1999), and the recent  compilation by Bensby et
al.  (2003)  are    reported in   Table~\ref{tablines}.  The   adopted
oscillator strengths for these lines were then selected  from a fit to
the  solar spectrum observed with  the  same VLT-UVES instrumentation,
adopting a NMARCS solar atmospheric  model (Edvardsson et al.   1993),
in order to be compatible with the model  grid used to  analyse the 
sample giants.
  These gf-values  were  checked  against the spectrum of
Arcturus  (Hinkle et al.  2000), by   adopting the stellar  parameters
derived  by Mel\'endez et  al. (2003), namely T$_{\rm  eff}$ = 4275 K,
log g = 1.55,  [Fe/H]  = -0.54, v$_{\rm  t}$ =  1.65 km s$^{-1}$   and
abundances [C/Fe] = -0.08,  [N/Fe] = +0.3,  [O/Fe] = +0.43,  and other
abundance ratios from the literature, and rederived from our fits:
[Na/Fe]  = +0.1, [Mg/Fe]  =  +0.35, [Al/Fe]  = +0.3, [Si/Fe]  = +0.25,
[Ca/Fe] = +0.1, [Ti/Fe] = +0.3.  The  selected gf-values are marked in
bold face in Table~\ref{tablines}.

For  the  oxygen  forbidden  line  [OI]6300 {\rm   \AA}  we adopt  the
oscillator strength recently derived by  Allende Prieto et al. (2001):
log gf = -9.716.

For lines  of  the heavy  elements BaII, LaII  and  EuII,  a hyperfine
structure was taken into account,  based on the hyperfine constants by
Lawler et al.  (2001a) for EuII,  Lawler et  al. (2001b) for  LaII and
Biehl (1976) for BaII.

Solar abundances were adopted from Grevesse  et al. (1996), except for
the  value  for oxygen  where  $\epsilon$(O)  = 8.77  was  assumed, as
recommended by Allende Prieto et  al. (2001) for the  use of 1-D model
atmospheres.


\section{Iron abundances}

Photospheric 1-D models for the sample giants  were extracted from the
NMARCS grid (Plez   et al.  1992),  originally  developed  by Bell  et
al. (1976) and Gustafsson et al. (1975).

The LTE  abundance analysis  and  the spectrum  synthesis calculations
were   performed  using  the  codes   by  Spite (1967, and  subsequent
improvements made in the last thirty years), Cayrel et al.  (1991) and
Barbuy et al. (2003).

The stellar   parameters  were  derived    by initially  adopting  the
photometric effective  temperature  and  gravity,  and   then  further
constraining  the  temperature by imposing  excitation equilibrium for
\ion{Fe}{I} lines and  the gravity by imposing  ionization equilibrium
for \ion{Fe}{I} and \ion{Fe}{II}.   Microturbulence velocity v$_t$ was
determined by canceling  the  trend of Fe  abundance   vs.  equivalent
width.
 
The same method   and  the same line   list,  applied to  the Sun  and
Arcturus,    yield    metallicities  of  [\ion{Fe}{I}/H]      =  0.09,
[\ion{Fe}{II}/H] =  0.01 and [\ion{Fe}{I}/H] = -0.49, [\ion{Fe}{II}/H]
= -0.56 respectively.  Note that  in this case the stellar  parameters
that we  converged to are  T$_{\rm eff}$ 5800 K, log  g = 4.4, v$_{\rm
t}$ = 1.0  km s$^{-1}$ for the  Sun and T$_{\rm eff}$ = 4300 K, log g =
1.8, v$_{\rm t}$  =  1.6 km s$^{-1}$ for  Arcturus,  in good agreement
with expectations.

\begin{figure}[ht]
\psfig{file=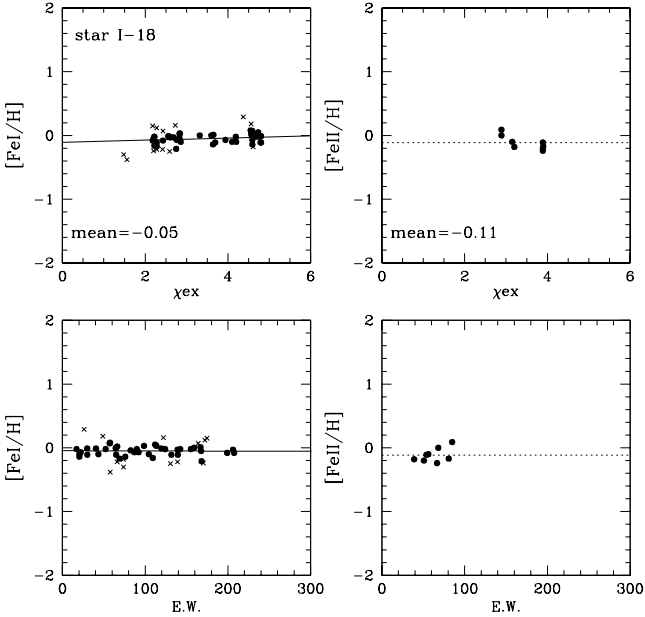,angle=0,width=9cm}
\caption {Iron abundance for each line measured in I-18.
Abundances of FeI and FeII are shown  as a function of both $\chi_{ex}$
and E.W. in the  different panels. The lines  shown with a cross  have
been excluded from   the computation of  the  mean abundance, using  a
$\sigma$ clipping criterion. The   flattening of the FeI  abundance  with
$\chi_{ex}$ and    E.W. has  been  used to    set constraints  on  the
temperature and microturbulence velocity, respectively.}
\label{I18feh}
\end{figure}

\begin{figure}[ht]
\psfig{file=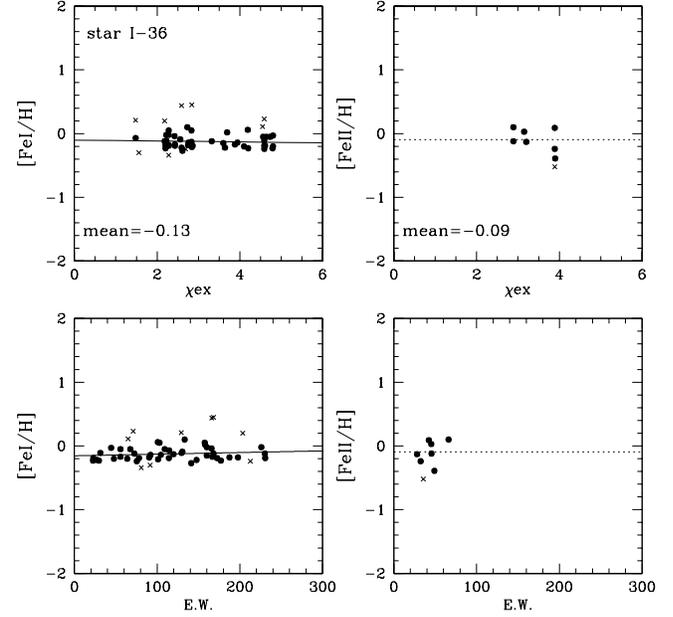,angle=0,width=9cm}
\caption {Same as Fig.~\ref{I18feh} for star I-36.}
\label{I36feh}
\end{figure}

\begin{figure}[ht]
\psfig{file=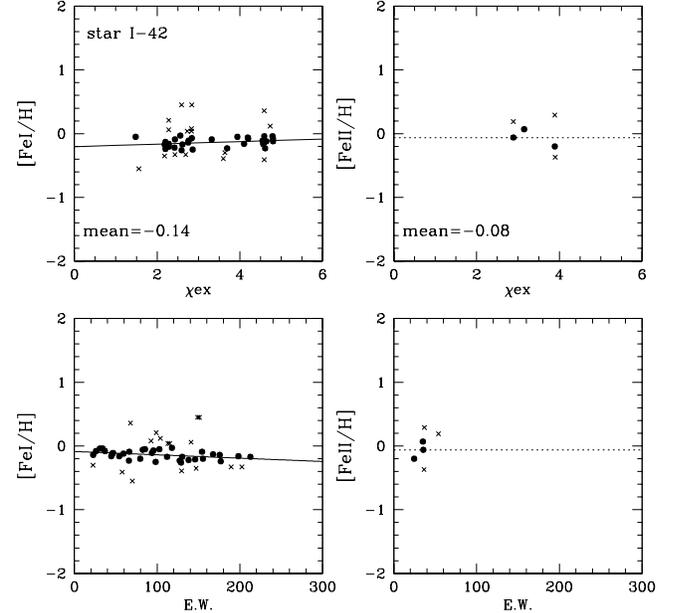,angle=0,width=9cm}
\caption {Same as Fig.~\ref{I18feh} for star I-42.}
\label{I42feh}
\end{figure}

Table~\ref{tabfeh} reports [\ion{Fe}{I}/H] and [\ion{Fe}{II}/H] values
for   the  sample    stars.    Figs.~\ref{I18feh}, \ref{I36feh}    and
\ref{I42feh} illustrate the  results of the  analysis for stars  I-18,
I-36 and I-42 respectively.   The  abundances from \ion{Fe}{II}  lines
for star  I-42  showed relatively  large scatter,  possibly  due  to a
combination of the lower   S/N  and cooler temperature.  We  therefore
compared the equivalent width of \ion{Fe}{II} lines in I-42 with those
of I-36, having very  similar T$_{\rm  eff}$,  and rejected the  lines
showing  large deviations from   the  mean  percent difference.   This
allowed  us   to have a  smaller    dispersion in [\ion{Fe}{II}/H] and
therefore to get a spectroscopic constraint on the gravity for I-42.

A second set  of  parameters based on  the  photometry was derived  by
imposing the temperature and  gravity values  from Table~\ref{tabteff}
and determining the metallicity and the microturbulence velocity from
\ion{Fe}{I} and \ion{Fe}{II}. The two sets of stellar parameters
are given in Table~\ref{tabfeh}.
 
As  a  check  ionization  equilibrium  was  also  used   as a
temperature  indicator,  as  suggested  by  McWilliam  \&  Rich (2003,
hereafter  MR03), keeping the  gravity fixed to its photometric value,
and     varying  the  temperature     to   recover the Fe   ionization
equilibrium. This leads  to  temperatures in better agreement  with the
photometric values, while   keeping excitation equilibrium   virtually
identical to the formally best fitting spectroscopic temperature.

This shows  that the true  temperature  of  the sample stars  is  very
likely hotter  (by $\sim  200$ K) than  the   photometric one. In  the
following we will keep this  last set of  parameters as the one  which
gives  the  most self consistent picture.  The   errors on  [Fe/H] were
estimated as follows. The  formal error on  the slope in the FeI $vs.$
ionization potential implies an error in the temperature of $\pm$100 K
for  all the  stars.  From  the  abundance analysis  of  Arcturus,  we
estimated that  a    variation of $\pm$    100 K  in  the  temperature
propagates into a difference of $\pm$0.07 in [FeI/H] and $\pm 0.06$ in
[FeII/H]. This variation  in the metallicity  of Arcturus was found by
forcing excitation equilibrium (i.e, by fixing the gravity) and canceling
the trend in abundance  with E.W.  (i.e., by fixing  $v_t$)  with the
new (uncertain by $\pm 100$ K) temperatures.   This procedure gives an
estimate  of  the  systematic  error   due to    a possible  uncertain
assumption  on the input parameters.   The statistical error, measured
from the spread in abundance obtained from different lines, is
0.018 $<$ $\Delta$[FeI/H] $<$ 0.026 for the three  stars. The final error,
being the quadratic sum of  the systematic and statistical components,
is quoted in Table~\ref{tabfeh}.

\begin{table}
\caption[6]{Stellar parameters and derived metallicities}
\begin{flushleft}
\tabcolsep 0.15cm
\begin{tabular}{lllllc}
\noalign{\smallskip}
\hline
\noalign{\smallskip}
{\rm star} &${\rm T_{eff}}$ & ${\rm log\; g}$ & ${\rm [FeI/H]}$ & ${\rm [FeII/H]}$ 
& ${\rm v_t (km/s)}$  \\
\noalign{\smallskip}
\hline
\noalign{\smallskip}
\multicolumn{4}{l}{Spectroscopic:}\\
{\rm I-18}  & 4800 & 2.3 &  +0.05 ($\pm0.072$)  & --0.07 ($\pm0.08$) & 1.5   \\
{\rm I-36}  & 4300 & 1.8 & --0.04 ($\pm0.073$)  & --0.08 ($\pm0.10$) & 1.5   \\
{\rm I-42}  & 4200 & 1.6 & --0.16 ($\pm0.075$)  & --                 & 1.3   \\
&&&&&\\
\multicolumn{4}{l}{Photometric:}\\
{\rm I-18}  & 4500 & 2.0 & --0.05 ($\pm0.072$)  &  +0.18 ($\pm0.08$) & 1.3   \\
{\rm I-36}  & 4000 & 1.5 &  +0.00 ($\pm0.073$)  &  +0.35 ($\pm0.10$) & 1.5   \\
{\rm I-42}  & 4000 & 1.5 & --0.19 ($\pm0.075$)  &  --                & 1.3   \\
&&&&&\\
\multicolumn{4}{l}{Ionization temperature:}\\
{\rm I-18}  & 4700 & 2.0 & --0.05 ($\pm0.072$)  & --0.11 ($\pm0.08$) & 1.5   \\
{\rm I-36}  & 4200 & 1.5 & --0.13 ($\pm0.073$)  & --0.09 ($\pm0.10$) & 1.5   \\
{\rm I-42}  & 4100 & 1.6 & --0.14 ($\pm0.073$)  & --0.08 ($\pm0.10$) & 1.2   \\
\noalign{\smallskip} 
\hline 
\end{tabular}
\label{tabfeh}
\end{flushleft} 
\end{table}

\section{Abundance ratios}

Elemental  abundances  were   obtained  through line-by-line  spectrum
synthesis calculations, for the line list given in Tables~\ref{tablines}
and \ref{tabCNO}.

The calculations of synthetic spectra were carried out as described in
Cayrel et al.  (1991) and Barbuy et al.  (2003), where molecular lines
of CN  (A$^2$$\Pi$-X$^2$$\Sigma$), C$_2$ Swan (A$^3$$\Pi$-X$^3$$\Pi$), TiO
(A$^3$$\Phi$-X$^3$$\Delta$) $\gamma$   and  TiO (B$^3$$\Pi$-X$^3$$\Delta$)
$\gamma$' systems  are taken   into  account.  As a first   guess, the
abundances obtained from   the equivalent width, in  the ``classical''
way,  without taking  into    account the effect of   molecules,  were
adopted. These first guess values are listed in Table~\ref{tabguess}.

\begin{figure}[ht]
\psfig{file=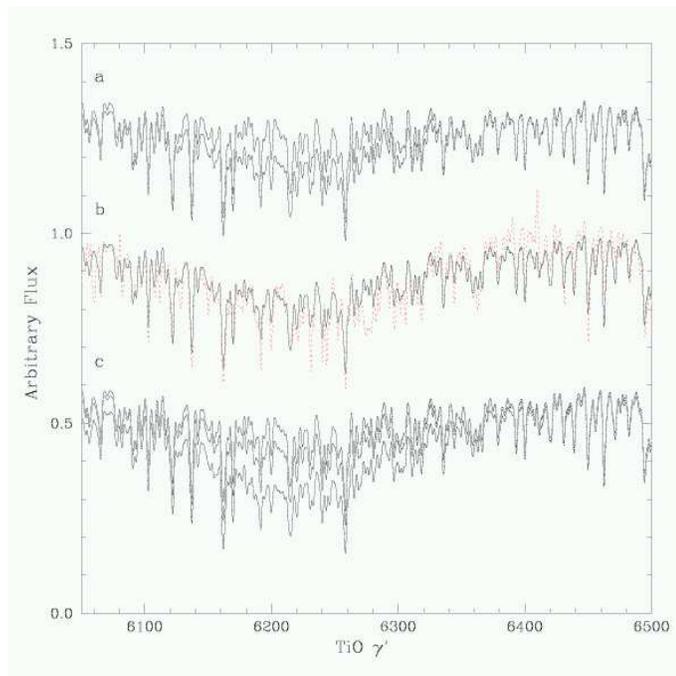,angle=0,width=9cm}
\caption {TiO $\gamma$' bandhead in star I-42, where spectra
are convolved with gaussians of FWHM = 2.0 {\rm \AA}:
a) computed with [O/Fe]=[Ti/Fe] = 0.0 and +0.2 (adopted value);
b) comparison of the observed spectrum (dotted line)
 with model computed with
T$_{\rm eff}$ = 4100 K and [O/Fe]=[Ti/Fe]=+0.2;
c) T$_{\rm eff}$ = 4000, 4100 and 4200 K.
}
\label{tio}
\end{figure}

The comparison between the synthetic spectrum, generated including the
effects  of  molecules,  and  the   observed one,  for  each line   in
Table~\ref{tablines}, allowed us to  set a more realistic estimate  of
the  true      abundance ratios.   Our best    values    are listed in
Table~\ref{tabalpha}.  The comparison between Table~\ref{tabguess} and
Table~\ref{tabalpha} gives an estimate of the importance of the effect
of molecule formation in the atmosphere of each star.  The main effect
is a lowering of the continuum as  illustrated in Fig~\ref{tio}, where
TiO was  computed for [O/Fe]=[Ti/Fe]$=0.0$  and +0.2 (the latter value
is the adopted one), and for T$_{\rm eff}=4000$, 4100 and 4200 K.

Errors in the abundance ratios can be due essentially to gf-values and
damping     constants    used.     They   can   be      deduced   from
Table~\ref{tablines}, where a  difference   of  0.2  dex in   log   gf
essentially means a  difference of 0.2 dex   in abundance, for a  given
damping constant.

The carbon abundance [C/Fe] = 0.0 was derived for  the 3 sample stars,
based on the C$_2$(0,1) 5635 {\rm \AA}, whereas the nitrogen abundance
[N/Fe] =   +0.6 was derived from  a  series of CN  bandheads along the
spectrum.  The region of the C$_2$  line is very  noisy, and the whole
region  is  blended  with CN  lines   as well, such    that the carbon
abundance is not as precise as all the other derivations in this work.
On the other hand, C+N is reliable. The N overabundance corresponds to
a mixing that is expected in giants, bringing carbon down and nitrogen
up,  but  since  C+N is   high,  it does possibly   involve  as well a
primordial enhancement in C+N elements, as detected before in NGC 6553
(Mel\'endez et al. 2003).

\begin{figure}[ht]
\psfig{file=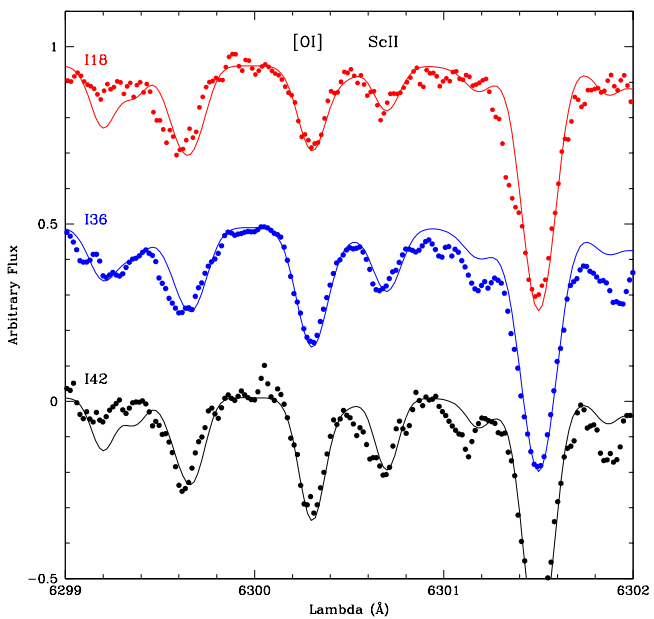,angle=0,width=9cm}
\caption {Fit of the [OI]6300.311 {\rm \AA} line in the program stars.}
\label{Oxy}
\end{figure}

\begin{figure}[ht]
\psfig{file=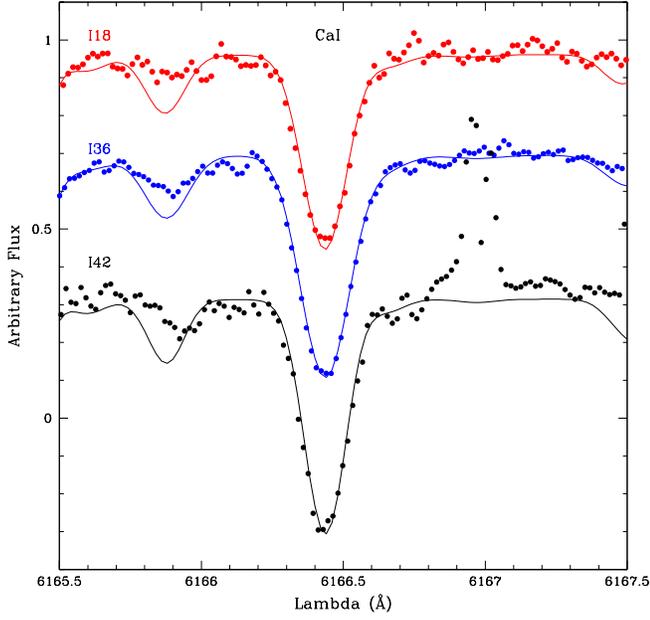,angle=0,width=9cm}
\caption {Fit of the CaI 6166.44 {\rm \AA} line in the program stars.}
\label{Calcium}
\end{figure}

\begin{figure}[ht]
\psfig{file=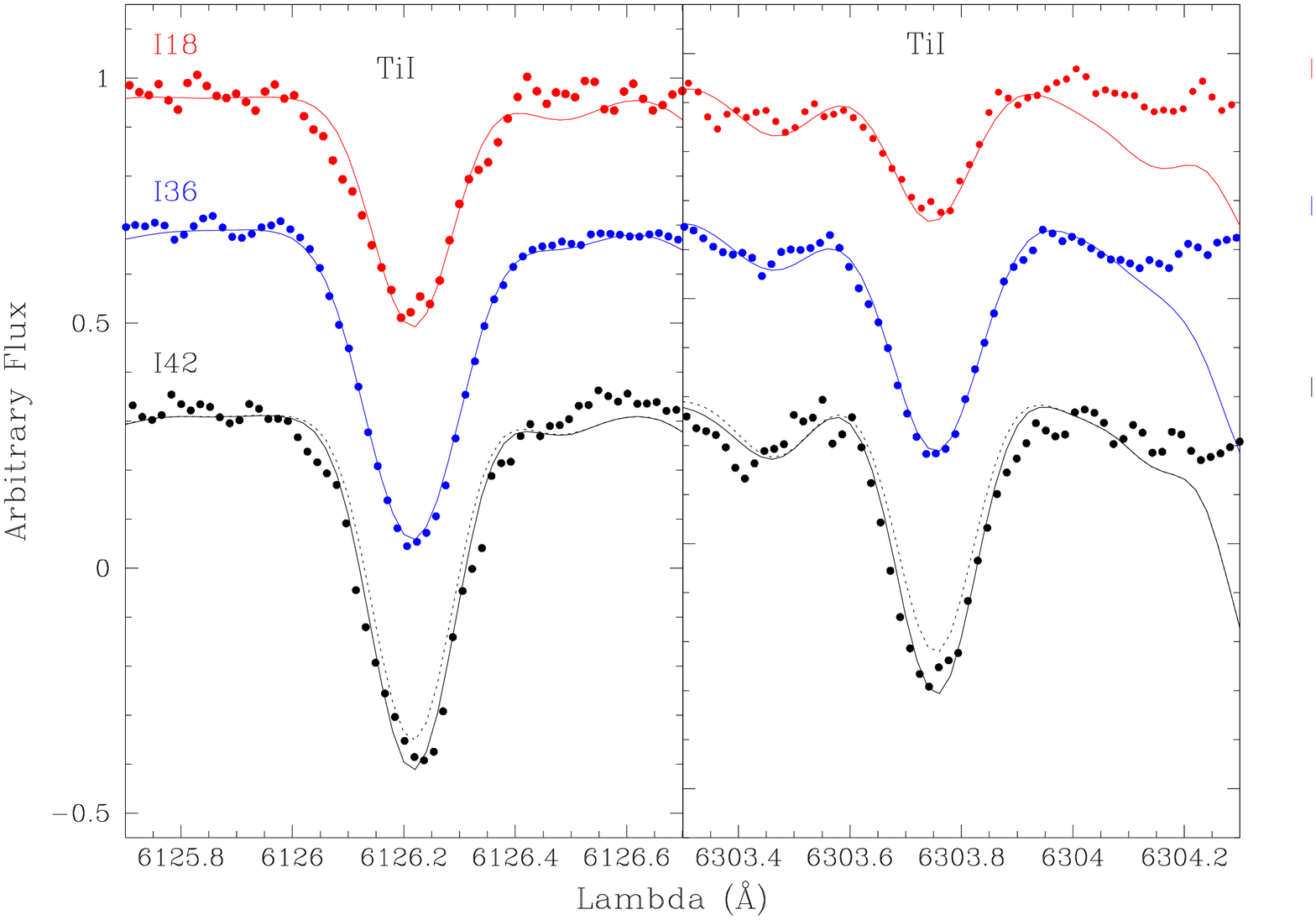,angle=-90,width=9cm}
\caption {Fit of the TiI 6126.224 and 6303.77 {\rm \AA} lines in the 
program stars.}
\label{Titanium}
\end{figure}

Figs.~\ref{Oxy},  \ref{Calcium}, \ref{Titanium} show  examples of  the
fit of Oxygen, Calcium and  Titanium lines, respectively, in the three
sample stars.  While   the abundance of   [O/Fe]  and [Ca/Fe] is  very
homogeneuos  in the three   stars, Titanium is considerably higher  in
star I-42. Fig.~\ref{Titanium} shows,  for comparison, also  a model
with low     Ti ([Ti/Fe]=-0.2)    with the  stellar      parameters of
I-42. Clearly such  low Ti for I-42 can  be excluded from the  present
data.

The fit  of   the  Oxygen line    [OI]6300.311  {\rm \AA}   shown   in
Fig.~\ref{Oxy} also shows the   nearby ScII 6300.67 line.  Sneden   \&
Primas  (2001) pointed out that  the ratio of  [OI] to ScII lines is a
more reliable measure of the  oxygen enhancement.  Indeed, OI and ScII
are  the  dominant species     of  these  elements in  cool    stellar
atmospheres, and  excitation  potentials of the [OI]6300.31  {\rm \AA}
($\chi_{ex}$ = 0.0 eV ) and that of ScII6300.67 {\rm \AA} ($\chi_{ex}$
= 1.5 eV)  are similar. We  used a  hyperfine  structure splitting and
corresponding log  gf   of  the ScII    line  as given   in Spite   et
al.  (1989). The present analysis  gives [O/Sc]=+0.35, +0.35 and +0.20
for I-18, I-36 and I-42, respectively.

The line-by-line  abundance ratios  are given in Table~\ref{tablines},
while the  final  resulting abundance ratios  for  the 3 sample stars,
together with their mean, are shown in Table~\ref{tabalpha}.  Column 7
lists the results for the 4 HB stars analysed by C01 and column 8 shows
the average of Baade's Window 11 K  giants analysed by MR94.  There is
a general good agreement with respect to  C01, except for a difference
of  0.2 dex in [FeI/H], and  the Si, Ca  and BaII abundances which are
higher in their study.  The Mg, Al and Ti abundances  found by MR94 in
Baade's Window are  higher than the  values found in the present paper
and in C01.

\begin{table*}
\caption[10]{Abundance ratios using the ``Ionization temperature'' set of
parameters given in Table\ref{tabfeh}. ($^{\rm a}$): C$_6$ computed using
collisional broadening theory by Barklem et al. (1998, 2000).}
\begin{flushleft}
\tabcolsep 0.15cm
\begin{tabular}{llllrrrrrrrrrrrrrrrr}
\noalign{\smallskip}
\hline
\noalign{\smallskip}
Species & $\lambda$ & $\chi_{ex}$ &  C$_6$ &\multispan6 loggf & \multispan4 [X/Fe]  \\
 &  &  &  & \multispan6  $---------------------$ & \multispan4  $----------$  \\
 & (${\rm \AA}$) & (eV)  &  & BW92 & MR94 & NIST & B99 & VALD & BFL03 & $\alpha$Boo &  I-18 & I-36 & I-42  \\
\noalign{\smallskip}
\hline
\noalign{\smallskip}
NaI & 6154.230 & 2.10 & 0.90E-31 	   & -1.53 & -1.57       & {\bf -1.56} & -1.56       & -1.56 & -1.57       &  0.00 & +0.65 & +0.40 & +0.30 & \\
NaI & 6160.753 & 2.10 & 0.30E-31 	   & -1.23 & -1.27       & {\bf -1.26} & -1.26       & -1.26 & -1.27       & +0.10 & +0.60 & +0.40 & +0.30 & \\
MgI & 6318.720 & 5.11 & 0.30E-30 	   & --    & --          & --          & {\bf -2.10} & -1.72 & --          & +0.35 & +0.00 & -0.05 & +0.05 & \\
MgI & 6319.242 & 5.11 & 0.30E-31 	   & --    & -2.22       & --          & {\bf -2.36} & -1.95 & --          & +0.35 & +0.15 & -0.05 & +0.05 & \\
MgI & 6319.490 & 5.11 & 0.30E-31 	   & --    & --          & --          & {\bf -2.80} & -2.43 & --          & +0.35 & +0.00 & -0.05 & +0.05 & \\
MgI & 6765.450 & 5.75 & 0.30E-31 	   & --    & {\bf -1.94} & --          & -1.94       & --..  & --          &  --   & +0.00 & +0.05 & +0.25 & \\
SiI & 5948.548 & 5.08 & 2.19E-30$^{\rm a}$ & -1.23 & --          &-1.24        & {\bf -1.17} & -1.23 & -1.13       & +0.35 & +0.10 & +0.10 & +0.05 & \\
SiI & 6142.494 & 5.62 & 0.30E-31 	   & -1.39 & -- 	 & --          & {\bf -1.50} & -0.92 & -1.50       & +0.30 & +0.05 & +0.10 & --    & \\
SiI & 6145.020 & 5.61 & 0.30E-31 	   & -1.32 & --          & --          & {\bf -1.45} & -0.82 & -1.46       & +0.25 & +0.05 & +0.10 & +0.05 & \\
SiI & 6155.142 & 5.62 & 0.30E-30 	   & -0.55 & --		 & --          & {\bf -0.85} & -0.40 & -0.72       & +0.30 & +0.10 & +0.10 & +0.15 & \\
SiI & 6237.328 & 5.61 & 0.30E-30 	   & -0.84 & {\bf -1.01} & --          & -1.15       & -0.53 & -1.05       & +0.20 & +0.05 & +0.10 & --    & \\
SiI & 6243.823 & 5.61 & 0.30E-31 	   & -1.15 & --          & --          & -2.32       & -0.77 & {\bf -1.30} & +0.30 & +0.05 & +0.10 & +0.05 & \\
SiI & 6414.987 & 5.87 & 0.30E-30 	   & -0.89 & --          & --          & {\bf -1.13} & -1.10 & --          & +0.35 & +0.20 & +0.20 & --    & \\
SiI & 6721.844 & 5.86 & 0.90E-30 	   & -0.94 & {\bf -1.17} & -0.94       & -0.94       & -1.49 & --          & +0.10 & +0.00 & +0.10 & +0.10 & \\
CaI & 6102.727 & 2.52 & 4.54E-31$^{\rm a}$ & -0.79 & --          & {\bf -0.79} & -0.24       & -0.96 & --          & +0.10 & -0.30 & -0.40 & -0.40 & \\
CaI & 6156.030 & 2.52 & 0.30E-31 	   & --    & --          & -2.20       & {\bf -2.39} & -2.50 & --          & +0.10 & -0.35 & -0.30 & -0.40 & \\
CaI & 6161.295 & 2.51 & 5.98E-31$^{\rm a}$ & -1.03 & --          & {\bf -1.02} & -1.25       & -1.29 & -1.26       & +0.10 & -0.35 & -0.40 & -0.40 & \\
CaI & 6162.167 & 2.52 & 5.95E-31$^{\rm a}$ & -0.80 & --          & {\bf -0.09} & -0.10       & -0.17 & --          & +0.10 & -0.30 & --    & -- & \\
CaI & 6166.440 & 2.52 & 5.95E-31$^{\rm a}$ & -1.14 & -1.14       & {\bf -0.90} & -1.07       & -1.16 & -1.17       & +0.10 & -0.30 & -0.40 & -0.30 & \\
CaI & 6169.044 & 2.52 & 5.95E-31$^{\rm a}$ & -0.80 & -0.80       & {\bf -0.54} & -0.71       & -0.81 & -0.84       & +0.10 & -0.20 & -0.40 & -0.40 & \\
CaI & 6169.564 & 2.52 & 5.95E-31$^{\rm a}$ & -0.48 & -0.48       & {\bf -0.27} & -0.27       & -0.93 & -0.63       & +0.10 & -0.40 & -0.40 & --    & \\
CaI & 6439.080 & 2.52 & 5.12E-32$^{\rm a}$ & +0.47 & --          & +0.47       & -0.04       & +0.39 & {\bf +0.3}  & +0.10 & -0.30 & -0.40 & -0.40 & \\
CaI & 6455.605 & 2.52 & 5.06E-31$^{\rm a}$ & --    & --          & {\bf -1.35} & -1.35       & -1.56 & -1.30       & +0.15 & -0.40 & -0.40 & -0.40 & \\
CaI & 6464.679 & 2.52 & 5.09E-32$^{\rm a}$ & --    & --          & --          & {\bf -2.10} & -4.27 & --          & +0.10 & --    &  0.00 & -0.60 & \\
CaI & 6471.668 & 2.52 & 5.09E-32$^{\rm a}$ & -0.69 & --          & {\bf -0.59} & -1.00       & -0.65 & -0.65       & +0.10 & -0.30 & -0.30 & --    & \\
CaI & 6493.788 & 2.52 & 5.05E-32$^{\rm a}$ & -0.11 & --          & {\bf +0.14} & +0.60       & +0.02 & -0.19       & +0.10 & -0.35 & -0.30 & -0.40 & \\
CaI & 6499.654 & 2.52 & 5.05E-32$^{\rm a}$ & -0.82 & --          & {\bf -0.59} & -0.63       & -0.72 & -0.73       & +0.10 & -0.30 & -0.30 & -0.40 & \\
CaI & 6508.846 & 2.52 & 0.30E-31 	   & -2.41 & {\bf -2.50} & -2.11       & -2.51       & -2.16 & --          & +0.15 & -0.20 & -0.40 & -0.20 & \\
CaI & 6572.779 & 0.00 & 2.62E-32$^{\rm a}$ & -4.31 & -4.31       & {\bf -4.29} & -4.30       & -4.10 & --          & +0.10 & -0.40 & -0.60 & -0.20 & \\
CaI & 6717.687 & 2.71 & 0.70E-30 	   & --    & --          & {\bf -0.61} & -0.61       & -0.60 & --          & +0.20 & -0.30 & -0.40 & -0.40 & \\
CaI & 6719.687 & 2.71 & 0.70E-30 	   & -2.41 & -2.55       & --          & {\bf -2.80} & --    & --          & +0.10 & --    & --    & --    & \\
TiI & 5866.452 & 1.07 & 2.16E-32$^{\rm a}$ & -0.84 & --          & {\bf -0.84} & -0.75       & -0.84 & --          & -0.40 & -0.20 & -0.20 & +0.10 & \\
TiI & 5922.110 & 1.05 & 3.46E-32$^{\rm a}$ & -1.47 & --          & {\bf -1.47} & -1.47       & -1.47 & --          & -0.40 & -0.30 & -0.20 & +0.20 & \\
TiI & 5965.835 & 1.88 & 2.14E-32$^{\rm a}$ & -0.41 & --          & {\bf -0.41} & -0.50       & -0.41 & --          & -0.40 & -0.30 & -0.30 & +0.10 & \\
TiI & 5978.549 & 1.87 & 2.14E-32$^{\rm a}$ & -0.50 & --          & {\bf -0.50} & -0.51       & -0.50 & --          & -0.30 & -0.40 & -0.20 & +0.10 & \\
TiI & 6064.626 & 1.05 & 2.06E-32$^{\rm a}$ & -1.94 & --          & {\bf -1.94} & -2.78       & -1.94 & --          & -0.40 & -0.30 & -0.20 & +0.20 & \\
TiI & 6091.177 & 2.27 & 3.89E-32$^{\rm a}$ & -0.42 & --          & {\bf -0.42} & -0.32       & -0.42 & --          & -0.30 & -0.30 & -0.20 & +0.10 & \\
TiI & 6126.224 & 1.07 & 2.06E-32$^{\rm a}$ & -1.42 & --          & {\bf -1.43} & -1.30       & -1.43 & -1.37       & -0.30 & -0.30 & -0.20 & +0.20 & \\
TiI & 6258.110 & 1.44 & 4.75E-32$^{\rm a}$ & -0.35 & --          & {\bf -0.36} & -0.51       & -0.36 & --          & -0.40 & -0.30 & -0.30 & +0.10 & \\
TiI & 6261.106 & 1.43 & 4.68E-32$^{\rm a}$ & -0.48 & --          & {\bf -0.48} & -0.66       & -0.48 & -0.42       & -0.50 & -0.35 & -0.30 & +0.20 & \\
TiI & 6303.767 & 1.44 & 1.53E-32$^{\rm a}$ & --    & -1.57       & {\bf -1.57} & -1.57       & -1.57 & -1.51       & -0.30 & -0.30 & -0.20 & +0.20 & \\
TiI & 6336.113 & 1.44 & 0.30E-31 	   & --    & -1.74       & {\bf -1.74} & -1.79       & -1.74 & --          & -0.30 & -0.20 & -0.20 & +0.20 & \\
TiI & 6508.154 & 1.43 & 1.46E-32$^{\rm a}$ & --    & {\bf -2.05} & --          & -2.20       & -2.15 & --          & --    & -0.30 & -0.20 & +0.20 & \\
TiI & 6554.238 & 1.44 & 2.72E-32$^{\rm a}$ & -1.22 & -1.22       & {\bf -1.22} & -1.22       & -1.22 & --          & -0.30 & -0.20 & -0.20 & +0.20 & \\
TiI & 6556.077 & 1.46 & 2.74E-32$^{\rm a}$ & -1.07 & -1.07       & {\bf -1.07} & -1.07       & -1.07 & --          & -0.35 & -0.05 & -0.30 & +0.20 & \\
TiI & 6599.113 & 0.90 & 2.94E-32$^{\rm a}$ & -2.09 & --          & {\bf -2.09} & -3.00       & -2.09 & --          & -0.35 & -0.30 & -0.20 & +0.10 & \\
TiI & 6743.127 & 0.90 & 0.30E-31 	   & --    & -1.63       & {\bf -1.63} & -1.68       & -1.63 & -1.63       & -0.40 & -0.35 & -0.30 & +0.10 & \\

TiII& 6491.580 & 2.06 & 0.30E-31 	   & --    & --          & {\bf -2.10} & -2.10       & -1.79 & --          & -0.30 & -0.30 & -0.20 & +0.10 & \\
TiII& 6559.576 & 2.05 & 0.30E-31 	   & --    & {\bf -2.48} & --          & -2.60       & -2.02 & --          & --    & -0.10 & -0.30 & --    & \\
TiII& 6606.970 & 2.06 & 0.30E-31 	   & --    & --          & {\bf -2.79} & -3.00       & -2.79 & -2.76       & -0.30 & -0.20 & -0.20 & +0.00 & \\
\noalign{\smallskip} \hline \end{tabular}
\label{tablines}
\end{flushleft} 
\end{table*}

\begin{table}
\caption[7]{C, N, O and heavy elements.
Carbon abundances are derived from C$_2$ lines and
nitrogen abundances from CN lines (Sect. 6).
Abundances of the heavy elements Ba, La and Eu are derived
using hyperfine structure (Sect. 4). }
\tabcolsep 0.13cm
\begin{tabular}{lcccrrrr}
\noalign{\smallskip}
\hline
\noalign{\smallskip}
Species & $\lambda ({\rm \AA})$ & $\chi_{ex}$(eV) & log gf &   I-18 &  I-36 & I-42 & \\
\hline
\noalign{\smallskip}
\hline
\noalign{\smallskip}
[OI]     & 6300.311 & 0.00 & --9.716 & +0.10 & +0.15 & +0.20  \cr
ScII     & 6300.67 & 1.50 & hfs & -0.25 & -0.20 & +0.00  \cr
C(C$_2$) & 5634.2,35.5 & (0,1) & --     &  0.00 &  0.00 &  0.00  \cr
N(CN)    & 6332.2 & (5,1) & --     & +0.60 & +0.60 & +0.60  \cr
BaII     & 6141.727 & 0.70 &    hfs & -0.20 & -0.15 & +0.10 \cr
BaII     & 6496.908 & 0.60 &    hfs & -0.15 & -0.15 & +0.10 \cr
LaII     & 6390.480 & 0.32 &    hfs &  0.00 & -0.15 & +0.20 \cr
EuII     & 6645.127 & 1.38 &    hfs &  0.00 & +0.10 & +0.10 \cr
\noalign{\smallskip} \hline \end{tabular}
\label{tabCNO}
\end{table}

\begin{table}
\caption[8]{{\it First guess} abundance ratios [X/Fe] of the program stars}
\tabcolsep 0.15cm
\begin{center}
\begin{tabular}{lcrrr}
\noalign{\smallskip}
\hline
\noalign{\smallskip}
Species & No. &   I-18 &  I-36 & I-42 \cr
\hline
\noalign{\smallskip}
\hline
\noalign{\smallskip}
[AlI/Fe]   & 2  &  -0.15 &  +0.00 &  +0.15 \cr
[NaI/Fe]   & 2  &  +0.62 &  +0.30 &  +0.18 \cr
[MgI/Fe]   & 4  &  +0.24 &  +0.15 &  +0.31 \cr
[SiI/Fe]   & 3  &  +0.12 &  +0.08 &  +0.11 \cr
[CaI/Fe]   & 7  & --0.30 & --0.41 & --0.28 \cr
[TiI/Fe]   & 6  & --0.30 & --0.08 &  +0.23 \cr
[TiII/Fe]  & 1  & --0.15 & --     &  +0.26 \cr
\noalign{\smallskip} \hline \end{tabular}
\label{tabguess}
\end{center}
\end{table}

\begin{table}
\caption[9]{Final abundance ratios [X/Fe] of the program stars}
\tabcolsep 0.13cm
\begin{tabular}{lcrrrrrrrrr}
\noalign{\smallskip}
\hline
\noalign{\smallskip}
Species & No. &   I-18 &  I-36 & I-42 & mean & C01 & MR94 \\
\hline
\noalign{\smallskip}
\hline
\noalign{\smallskip}
[FeI/H]    & 50 & --0.05 & --0.13 & --0.14 & --0.11  & +0.07  & --    \cr
[OI/Fe]    & 1  &  +0.10 &  +0.15 &  +0.20 &  +0.15  & +0.07  & +0.03 \cr
[OI/Sc]     & 1  &  +0.35 &  +0.35 &  +0.20 &  +0.30  & +0.17  & --    \cr
[NaI/Fe]   & 2  &  +0.60 &  +0.40 &  +0.30 &  +0.43  & +0.40  & +0.20 \cr
[MgI/Fe]   & 4  &  +0.05 &  +0.05 &  +0.10 &  +0.07  & +0.14  & +0.35 \cr
[SiI/Fe]   & 3  &  +0.05 &  +0.10 &  +0.10 &  +0.08  & +0.36  & +0.14 \cr
[CaI/Fe]   & 7  & --0.40 & --0.40 & --0.40 & --0.40  & +0.23  & +0.14 \cr
[TiI/Fe]   & 6  & --0.30 & --0.20 &  +0.20 & --0.10? & +0.03  & +0.37 \cr
[TiII/Fe]  & 1  & --0.20 & --0.20 &  +0.05 & --0.12? & --     & --    \cr
\noalign{\smallskip} \hline \end{tabular}
\label{tabalpha}
\end{table}


\section{Discussion and Conclusions}

We have derived element abundances  for  three giants in the  globular
cluster NGC~6528.   Five targets  were originally  selected.  However,
the high spatial resolution of  the HST/NICMOS imaging frames, and the
high  UVES spectral resolution  showed that two of  them are in fact a
blend of two stars (I-24 is likely a physical binary), and
for this reason the two objects have been excluded from the analysis.

This  case  shows how  strong can  be   the  effect of  projection or
physical binaries in similar   studies, when high  spatial  resolution
images of the field are not available.

From the   abundance analysis we   conclude that Iron is  about solar:
[Fe/H]$\sim -0.1$.

The odd-Z element  sodium,
built up     during  carbon   burning,   is  strongly     overabundant
([Na/Fe]$\sim$ 0.4).

The  $\alpha$-elements show a  quite peculiar behaviour.  
O, Mg and Si are just slightly above the solar value
([O/Fe] $\sim$[Mg/Fe]$\sim$  [Si/Fe]$\sim$0.1) while Ca  and  Ti are 
underabundant  with    respect   to the    Sun    ([Ca/Fe]$\sim -0.40$
[Ti/Fe]$\sim -0.2$). It is also somehow puzzling that despite the very
good agreement between the abundances  derived from different lines in
the  same  star (Table \ref{tablines}), the  three  stars show a
  spread  in  [Ti/Fe].   In  particular,    star  I-42   shows a
 higher Ti abundance with respect to the other two. Given
its  radial  velocity,  the similarity  with   I-18  and I-36  in  the
abundances of all  the other elements,  and the agreement between  its
photometric and spectroscopic  temperature,   we tend to discard   the
hypothesis of it being a field star just accidentally projected on the
cluster RGB sequence. We rather believe that its low temperature makes
stronger the effect of the TiO molecules in its atmosphere, which also
increases our uncertainty in the abundance analysis.

The  comparison between   our values   and   the ones  in Carretta  et
al. (2001) reveals  some discrepancies in   Si, Ca, and Ti. While  the
different treatment of  molecules in the two  analyses (Carretta et al.
did not account  for them) might explain part  of the discrepancy, the
large (0.6 dex) difference in the Ca  abundance has to arise from some
other factor. From an independent  analysis of the equivalenth  widths
of the NGC6528 stars observed by Carretta et al. (2001), MR03
found [Ca/Fe]=+0.2,  much  closer  to the  value found
here. Comparing the three  analyses, it seems  that the most  probable
source of discrepancy is the  adopted broadening factors. Indeed,  it
must be noted that all the CaI lines are rather strong, in the saturation
regime of the  curve  of growth.

In fact the abundances of Ca and Ti do not always follow the other 
$\alpha$-elements, as is the case for example with 47 Tucanae
where Ca-to-Fe is solar, and the other $\alpha$s are overabundant
(Brown \& Wallerstein 1992; Alves-Brito et al. 2004). According to
Woosley (private communication), there is metallicity sensitivity
in the yields because the mass loss varies with metallicity.
The yields therefore strongly depend on the mass loss, such that
Ca and Ti show variable proportions in different sites (Woosley
\& Weaver 1995), and [Ti/Si] is often low.

From our analysis we find a modest enhancement of  O, Mg and Si, which
is  consistent with the trend of  a lower $\alpha$-enhancement for the
metal-rich end  of the bulge  field  stars found  by McWilliam \& Rich
(2003).   However, the predictions by Matteucci  \& Brocato (1990) and
Matteucci, Romano and Molaro (1999)  give higher expected values for O
and   Mg ([O/Fe] $\sim$  [Mg/Fe]  $\sim$  +0.6  at  [Fe/H] =  0.0, and
somewhat less for Si and Ca ([Si/Fe]  $\sim$ [Ca/Fe] $\sim$ +0.3).  On
the other hand, the low values of  [Ca/Fe] $\approx$ [Ti/Fe] $\approx$
[Eu/Fe] $\approx$ 0.0 might indicate  that there has been a deficiency
of SNII of low mass, as pointed out in MR03,
and according to predictions by Woosley \& Weaver (1995). 
It is also worth noticing
that   Ca underabundance is   very    common in elliptical    galaxies
(e.g. Thomas et al. 2003 for a discussion of possible causes).


The  s-process elements Ba, La tend  to show  underabundances which is
expected  for an old   population.   The s-processing  takes  place in
massive stars (Raiteri  et al. 1993), as well  as in  relatively short
lived and long  lived intermediate mass  stars (Gallino et al.  1998),
indicating that  there has been no  significant contribution from long
lived stars.


\begin{acknowledgements}

We thank  an  anonymous referee  for his  comments, that improved  the
readability  of the manuscript.  We  are grateful  to Stan Woosley for
helpful comments on nucleosynthesis and chemical evolution of $\alpha$
elements, and to Cristina Chiappini for comments on chemical evolution
predictions.  BB acknowledges     grants  from CNPq  and  Fapesp.   DM
acknowledges grants from FONDAP Center for Astrophysics 15010003.

\end{acknowledgements}


\end{document}